\documentclass{article}

\usepackage[final]{neurips_2022}

\usepackage{verbatim}

\usepackage[utf8]{inputenc} 
\usepackage[T1]{fontenc}    
\usepackage{hyperref}       
\usepackage{url}            
\usepackage{booktabs}       
\usepackage{amsfonts}       
\usepackage{amsmath}
\usepackage{mathrsfs}       
\usepackage{amsthm}
\usepackage{nicefrac}       
\usepackage{microtype}      
\usepackage{xcolor}         
\usepackage{graphicx}
\usepackage{mathtools}
\usepackage{todonotes}
\usepackage{subcaption}
\usepackage{enumitem} 
\usepackage{algorithm,algorithmic}
\usepackage{cleveref}
\usepackage[english]{babel}
\usepackage{thmtools} 
\usepackage{thm-restate}
\usepackage{multibib}
\usepackage[normalem]{ulem}

\definecolor{RoyalBlue}{rgb}{0.25, 0.41, 0.88}
\hypersetup{colorlinks,linkcolor=purple,urlcolor=RoyalBlue,citecolor=purple}

\newcommand{\robin}[1]{\textcolor{olive}{[RG: #1]}}

\newcommand{\defeq}{\coloneqq}

\newcommand{\ul}{\underline}

\newcommand{\sas}{\sigma^2}
\newcommand{\ep}{\epsilon}

\newcommand{\sdp}{s_{\mathrm{dp}}}

\newcommand{\twid}[1]{\widetilde {#1}}

\newcommand{\iid}{\overset{\text{i.i.d.}}{\sim}}

\theoremstyle{plain}
\newtheorem{thm}{Theorem}[section] 

\newtheorem{assum}{Assumption}

\theoremstyle{definition}
\newtheorem{definition}[thm]{Definition}

\title{Data Augmentation MCMC for\\Bayesian Inference from Privatized Data}

\author{%
  Nianqiao Phyllis Ju\\
  Department of Statistics\\
  Purdue University\\
  West Lafayette, IN 47907 \\
  \texttt{nianqiao@purdue.edu} \\
   \And
   Jordan A. Awan \\
   Department of Statistics \\
   Purdue University\\
   West Lafayette, IN 47907\\
   \texttt{jawan@purdue.edu} \\
   \AND
   Ruobin Gong \\
   Department of Statistics \\
   Rutgers University \\
   Piscataway, NJ 08854 \\
   \texttt{ruobin.gong@rutgers.edu} \\
   \And
   Vinayak A. Rao \\
   Department of Statistics \\
   Purdue University\\
   West Lafayette, IN 47907\\
   \texttt{varao@purdue.edu} \\
}

\begin{document}

\maketitle

\begin{abstract}
Differentially private mechanisms protect privacy by introducing additional randomness into the data.
Restricting access to only the privatized data makes it challenging to perform
valid statistical inference on parameters underlying the confidential data.
Specifically, the likelihood function of the privatized data requires integrating over the large space of confidential databases and is typically intractable. 
For Bayesian analysis, this results in a posterior distribution that is doubly intractable, rendering traditional MCMC techniques inapplicable.
%
%
We propose an MCMC framework to perform Bayesian inference from the privatized data, which is applicable to a wide range of statistical models and privacy mechanisms. 
%
Our MCMC algorithm augments the model parameters with the unobserved confidential data, and alternately updates each one conditional on the other.
%
For the potentially challenging step of updating the confidential data, we propose a generic approach that exploits the privacy guarantee of the mechanism to ensure efficiency. 
We give results on the computational complexity, acceptance rate, and mixing properties of our MCMC. 
We illustrate the efficacy and applicability of our methods on a na\"ive-Bayes log-linear model and on a linear regression model.

\end{abstract}

\section{Introduction}\label{sec:intro}

{\bf Motivation.} Differential privacy \citep{dwork2006calibrating} presents a formal mathematical framework to protect the confidentiality of individuals and businesses in aggregate data products. It is the state-of-the-art standard for statistical disclosure limitation (SDL), and has become widely adopted by curators of large-scale scientific, commercial, and official databases. Differentially private data products are produced by probabilistic mechanisms that carry proven  privacy guarantees. Generally speaking, these mechanisms work by introducing carefully designed random noise into the query of interest, which is an otherwise deterministic function of the underlying database. 

The privatization of data products through noise infusion poses a challenge to statistical analysis in the downstream. Statistical estimators 
are typically complex functions of the data. If instead of the confidential data, 
the analyst only has access to a probabilistically processed version of them, how can they maintain the statistical validity of the resulting inference?
   
A crucial statistical advantage of differentially private mechanisms over traditional SDL counterparts, such as swapping \citep{Dalenius1977}, is that their probabilistic design is publicly known. This knowledge 
allows the data analyst to, at least in theory, accurately account for the privatization mechanism and conduct reliable uncertainty quantification. Nevertheless, it remains a substantial computational challenge to incorporate the privacy procedure into the statistical analysis. The challenge is a wide-spread and varied one, as the extra layer of privacy protection calls for the revision of a wide range of existing statistical methodologies that previously operate on the original, non-privatized data, most of which are neither low-dimensional nor simply structured. This is the challenge we address in this work, in which we develop a general computational framework for practitioners to obtain valid statistical inference based on privatized data. 

{\bf Related literature.} Current inferential strategies for privatized data fall into two broad categories. 
One invokes traditional statistical asymptotics to approximate the sampling distribution of a differentially private statistic, on the grounds that the privacy noise is often asymptotically negligible compared to errors due to sampling \cite[e.g.][]{smith2011privacy,cai2019cost}. These approximations are often inaccurate for finite sample sizes \cite[]{wang2018statistical} and call for specific handling to incorporate the privacy mechanism \cite[e.g.][]{gaboardi2016differentially,wang2015revisiting,gaboardi2018local}.

The second category recognizes (as Section~\ref{sec:marginal} will explain) that the marginal likelihood of the model parameters in~\eqref{eq:marginal} is central to the problem of inference from privatized data.
The marginal likelihood/marginal posterior distribution
requires a potentially high-dimensional integral over the space of unobserved confidential databases, and one that is analytically tractable only in a few, simple settings [e.g., \citealp{Awan2018:Binomial,awan2020differentially}].
Typically, one must resort to either approximating it or sampling from it using Monte Carlo methods.
Markov chain Monte Carlo (MCMC) techniques have been proposed for specific privacy mechanisms and data generating models. \citet{karwa2017sharing} propose an MCMC procedure for inference on exponential random graph models,
\cite{bernstein2018differentially,bernstein2019differentially} devise MCMC methods designed to handle the low-dimensional latent sufficient statistics from exponential family models and linear regression, and \cite{pmlr-v97-schein19a} design an MCMC algorithm for the Poisson factorization model when data are locally privatized under the double geometric mechanism. \cite{gong2019exact} shows that for certain differentially private statistics, approximate Bayesian computation (ABC) can give samples that are exact with respect to the marginal likelihood and the Bayesian posterior. In addition, {when the statistical model for the confidential data is fully parametric,}
the parametric bootstrap may be used to produce  inference accompanied by uncertainty quantification with better accuracy than asymptotic approximation \cite[e.g.][]{gaboardi2016differentially,ferrando2022parametric}. Variational Bayesian analysis \citep{Karwa2015:PrivatePD} is another alternative which invokes a non-asymptotic approximation to the posterior distribution.
These solutions are either approximations or hold only for specific settings.

{\bf Our contribution.}
We develop a general-purpose MCMC framework to perform Bayesian inference on the model parameters underlying the privatized data.
Our framework allows us to overcome the intractable marginal likelihood resulting from privatization, and is applicable to a wide range of statistical models and privacy mechanisms. The resulting MCMC algorithms are \emph{exact}, in that they target the posterior distribution precisely, without involving any approximation.

Our approach allows data analysts to leverage existing
inferential tools designed for non-private data. 
It can be viewed as a flexible, user-friendly wrapper that migrates existing MCMC algorithms for non-private
data to the setting of privatized data access, requiring no further algorithm design or tuning. 
The sampler, formally a Metropolis-within-Gibbs sampler, is presented in Algorithm~\ref{algo:ims}. 
It is general-purpose, requiring only that the analyst can 1) sample from the statistical model for the confidential data
and 2) can evaluate the probability density of the noise induced by the privacy mechanism. 
The algorithm augments the model parameters with the unobserved confidential data, and alternately updates each one conditioned on the other.
While the imputation of an entire unobserved database might appear daunting, we demonstrate how knowledge of the privacy mechanism can be exploited to confer performance guarantees to the proposed MCMC algorithm. We provide theoretical results for the computational complexity, Metropolis-Hastings acceptance rate, and mixing properties. 
In particular, the higher the privacy, the more rapid is our algorithm's exploration of the parameter space. 
We also identify a common structure present in many popular privacy mechanisms, which we term \emph{record additivity}. The record additivity of a privacy mechanism ensures that each round of our sampler has the same order of run-time as that of many non-private samplers. 
We illustrate the efficacy and applicability of our methods on a privatized na\"ive-Bayes log-linear model and a linear regression model with clamped and privatized input.
Source code in R are available at \url{https://github.com/nianqiaoju/dataaugmentation-mcmc-differentialprivacy}.

\section{Problem Setup}\label{sec:marginal}
Let $x = (x_1,\ldots,x_n) \in \mathbb{X}^n$ denote the confidential database, containing $n$ records. 
We assume these records are independent and identically distributed (i.i.d.) draws from a statistical model $f\left(\cdot \mid \theta\right)$, though this can be relaxed.
The goal of the analyst is to conduct statistical inference on the unknown model parameter $\theta\in \Theta$.
A Bayesian analyst represents {\em a priori} beliefs about $\theta$ with a prior probability distribution $p\left(\theta\right)$, and seeks to compute a posterior distribution $p(\theta \mid x) \propto p(\theta) f(x \mid \theta)$
that updates their beliefs in light of the observations $x$.
In many modern applications, this posterior distribution is intractable, and it is common for analysts to represent it using samples drawn via some MCMC algorithm. 
In this work, we will assume 
access to such a posterior sampling method:
\begin{assum}\label{assm:nonprivate}
The analyst has available a Markov kernel that targets $p(\theta \mid x) \propto p(\theta) f(x \mid \theta)$, the posterior distribution over the model parameters given the confidential database $x$.
\end{assum}



{\bf Differential privacy.}
Our work here focuses on the following departure from the usual Bayesian setting: instead of observing the database $x$, we observe a privatized data product or query, denoted as $\sdp$. The quantity $s_{dp}$  is typically of much lower dimensionality that the database $x$, and is probabilistically generated based on data $x$ through a \emph{privacy mechanism}, written as $\eta\left(\cdot \mid x\right)$. The privacy mechanism $\eta$ is said to be $\epsilon$-\emph{differentially private} ($\ep$-DP) \citep{dwork2006calibrating} if for  all values of $\sdp$, and for all `neighboring' databases $(x, x') \in  \mathbb{X}^n\times \mathbb{X}^n$ differing by one record 
(denoted by $d(x,x')\leq 1$), the probability ratio is bounded:
\begin{equation}\label{eq:eta-ratio}
    \frac{ \eta\left(\sdp \mid x\right)}{ \eta\left(\sdp  \mid x'\right)} \le \exp\left(\epsilon\right), \quad \epsilon > 0.
\end{equation}
The parameter $\epsilon$ is called the \emph{privacy loss budget}, and controls how informative $\sdp$ is about $x$. Large values of $\epsilon$ guarantee less privacy, while $\epsilon = 0$ corresponds to perfect privacy. 
A simple and widely used $\epsilon$-differentially private mechanism is the \emph{Laplace mechanism}: for a deterministic query $s:  \mathbb{X}^n \to \mathbb{R}^m$, the privatized query is defined as
$\sdp	=  s\left(x\right) + u$, 
where $u = \left(u_1,\ldots,u_m\right)$ are i.i.d.\ Laplace variables.
The scale parameter of the Laplace distribution is inversely proportional to $\epsilon$ (more privacy requires more noise), and directly proportional to
 $\Delta\left(s\right) = \max_{(x, x') \in  \mathbb{X}^n\times \mathbb{X}^n; d(x, x') \le 1} \left\Vert s\left(x\right) - s\left(x'\right)\right\Vert_{1}$,     
 the $\ell_1$ \emph{(global) sensitivity} of $s$ (the 
 more sensitive the confidential query is to changes in one record of the database, the more noise we need).  

Our methodology requires that the privacy mechanism $\eta$ is known and can be evaluated. 
This is true of $\epsilon$- (or \emph{pure}) DP, as well as common variants such as $(\epsilon,\delta)$- (or \emph{approximate}) DP, \emph{zero-concentrated} DP (zCDP)~\citep{dwork2016concentrated,bun2016concentrated}, and \emph{Gaussian}-DP \citep{dong2021gaussian}. 
To ensure computational efficiency, we make the following additional assumption. 


\begin{assum}[Record Additivity]\label{assm:rec_add}
The privacy mechanism can be written in the form $\eta(\sdp \mid x) = g\left(\sdp, \sum_{i=1}^n t_i(x_i,\sdp)\right)$ for some known and tractable functions $g, t_1,\dotsc,t_n$. 
\end{assum}

We refer to privacy mechanisms that satisfy~\Cref{assm:rec_add} as \emph{record-additive}. An implication of record additivity is that after changing one record in $x$, we do not have to scan the entire database to reevaluate $\eta$ . 
This is satisfied by many commonly used mechanisms, two important examples being: 1) mechanisms that add data-independent noise to a query of the form $s=\sum_{i=1}^n s_i(x_i)$, such as the sample mean, sample variance-covariance, and sufficient statistics of an exponential family distribution (see Sections~\ref{sec:loglinear} and~\ref{sec:regression} for examples), and  2) mechanisms designed to optimize empirical risk functions of the form $u(x,s_{dp}) = \sum_{i=1}^n u_i(x_i,s_{dp})$, such as the exponential mechanism \cite[]{mcsherry2007mechanism}, $K$-norm gradient mechanism \cite[]{reimherr2019kng}, objective perturbation \cite[]{Chaudhuri2011,Kifer2012:PrivateCERM}, and functional mechanism \cite[]{Zhang2012}.


{\bf Doubly intractable Bayesian inference from privatized data.}
Without access to the confidential database $x$, 
and given only the privatized query $\sdp$, the Bayesian analyst is now concerned with the following posterior distribution: 
\begin{equation}\label{eq:posterior}
    p\left(\theta \mid \sdp \right) \propto p\left(\theta\right)p\left(\sdp \mid \theta\right).
\end{equation}
Here, $p(\sdp|\theta)$ is the \emph{marginal likelihood} of $\theta$, integrating over all possible confidential databases: 
\begin{equation}\label{eq:marginal}
p\left(\sdp \mid \theta\right)=\int_{\mathbb{X}^n} \eta\left(\sdp \mid x\right) f\left(x\mid \theta\right) dx.
\end{equation}

The marginal likelihood contributes all the information that is available in the  privatized  observation $\sdp$ about the  parameter $\theta$, and is the foundation to statistical inference using privatized statistics \citep{williams2010probabilistic}. The posterior distribution~\eqref{eq:posterior} reveals that the inferential uncertainty about the parameter $\theta$ consists of three contributing sources: 1)  prior uncertainty as encoded in $p(\theta)$, 2) sampling (or modeling) uncertainty of the confidential database as reflected in $f$, and 3) uncertainty due to privacy as induced by the probabilistic mechanism $\eta$.

We now come to the core challenge to address in this work: the marginal likelihood in~\eqref{eq:marginal} calls for an integral over the entire space of possible input databases $x\in \mathbb{X}^n$. This is usually computationally challenging, especially if the privacy mechanism is not a function of a low-dimensional sufficient statistic. 
If the integral underlying the marginal likelihood is intractable, then $p(\sdp \mid \theta)$ cannot be analytically evaluated.
This makes the corresponding posterior distribution $p(\theta \mid \sdp)$ of~(\ref{eq:posterior}) \emph{doubly} intractable~\citep{murray2012mcmc} in the sense that it cannot be analytically evaluated even up to a normalizing constant.
Thus, traditional MCMC techniques are inapplicable and
inference strategies devised for privatized statistics must
tame this possibly high-dimensional integration problem. 

\section{Data Augmentation MCMC for Inference from Privatized Data}\label{sec:ourSampler}
In this paper, we present 
a simple, efficient, and general {\em data augmentation} MCMC~\citep{tanner1987calculation,van2001art} framework, allowing practitioners
to perform valid Bayesian inference on a wide-range of data models and privacy mechanisms.
Our approach is to augment the MCMC state space with the latent confidential database $x$, so that the stationary distribution is the \emph{joint} posterior distribution
\begin{equation}\label{eq:joint-posterior}
 p(\theta, x \mid \sdp)  \propto p(\theta) f(x \mid \theta) \eta(\sdp \mid x).
\end{equation}
Marginally, the $\theta$ samples produced by such an algorithm follow the posterior $p(\theta \mid \sdp)$ in \eqref{eq:posterior}. Our sampler is {\em exact}, targeting the marginal posterior distribution $p(\theta|\sdp)$ without any approximation error, despite the fact that the marginal likelihood \eqref{eq:marginal} is intractable.

Our approach of imputing the latent confidential database $x$ is motivated by two factors: 1) 
 we wish our algorithm to be {\em general-purpose}, applicable to a wide range of models and privacy mechanisms, and 2) we wish our algorithm to inherit guarantees on mixing performance from guarantees of the privacy mechanism. 
Towards these ends, we do not assume any specific form of 
the underlying model of $x$ and the privacy mechanism beyond Assumptions~\ref{assm:nonprivate} and~\ref{assm:rec_add} respectively. 
In this light, our contribution can be viewed as a flexible wrapper that allows existing MCMC algorithms for models of the confidential data to be extended to settings where the data is now protected by some privacy mechanism. 
Though imputing the confidential dataset might appear to present a significant challenge, we show that properties of the mechanism can be exploited to give performance guarantees on our sampling scheme, and show that it has a runtime of the same order as the non-private sampler.

In what follows, we outline our proposed MCMC algorithm, derive guarantees on the runtime and acceptance rate of the algorithm, and provide mild conditions for the proposed samplers to be ergodic, as well as additional conditions for 
our sampler to achieve geometric rates of convergence.

\subsection{A Privacy-Aware Metropolis-within-Gibbs Sampler} \label{sec:gibbs-overview}

Our approach to sample from the joint posterior distribution $p(\theta,x \mid \sdp)$ is through a sequence of alternating Gibbs updates. Let $(x^{(t)},\theta^{(t)})$ denote the state of the Gibbs sampler at the $t$-th iterations. 
Each iteration of the Gibbs sampler entails two steps:\\ 
%
({\bf Step 1}) sample $\theta^{(t+1)}$ from $p(\cdot \mid x^{(t)}, \sdp)$, and 
({\bf Step 2}) sample $x^{(t+1)}$ from $p(\cdot \mid \theta^{(t+1)},\sdp)$. 

The conditional distribution in Step 1 simplifies as $p(\theta|x^{(t)},\sdp) = p(\theta|x^{(t)})$,  highlighting why data-augmentation is useful: this 
conditional distribution is independent of the privacy mechanism, and we can use existing sampling algorithms (\Cref{assm:nonprivate}) for the confidential data. 
We note that with the exception of a few models, such as simple models with conjugate priors, it is usually not possible to directly sample from $p(\theta|x)$.
\Cref{assm:nonprivate} however only requires that we can {\em conditionally} simulate a new value of $\theta$ from a Markov kernel that has $p(\theta|x^{(t)})$ as its stationary distribution. 
Our overall Gibbs sampler then becomes a Metropolis-within-Gibbs sampler~\citep{gilks1995adaptive}, that nevertheless targets the joint posterior $p(\theta,x|\sdp)$.

Step 2 is the data-augmentation step, and 
connects the statistical model and the privacy mechanism on $x$.
Again, we cannot expect to produce a conditionally independent sample of the latent database from $p(x\mid \theta,\sdp)$, as this is model and mechanism dependent.
Instead, we take the much more tractable approach of cycling through the elements of latent database $x$, 
sequentially updating $x$ one element of a time.
Writing $x_{-i} = (x_1,\ldots,x_{i-1},x_{i+1},\ldots,x_n)$ to denote the vector $x$ excluding the $i$th element, 
step 2 then consists of the sequence of updates 
$p(x_1 \mid \theta, x_{-1}, \sdp), p(x_2 \mid \theta, x_{-2}, \sdp),\ldots, p(x_n \mid \theta, x_{-n}, \sdp)$. 
The complete sweep can be viewed as a dependent update of the latent database $x$ that targets the conditional distribution $p(x \mid \theta,\sdp)$.

Before we specify our complete sampler in Algorithm \ref{algo:ims}, we first  
address the following questions: 
(Q1)
{\em What is the performance loss from updating $x$ one element at a time, rather than jointly?}, and (Q2)
{\em How can we efficiently carry out the conditional updates $p(x_i \mid \theta, x_{-i},\sdp),\ i =1,\dotsc,n$?} 

Q1 concerns whether the dependence of $x_i$ given $(x_{-i},\sdp)$ is so strong as to impede efficient exploration of the $\mathbb{X}^n$-space and cause poor mixing. Here we note that the privacy mechanism limits the change in the likelihood $\eta(\sdp\mid x)$ when one element of $x$ is changed, and therefore limits the coupling between $x_i$ and $x_{-i}$.
This suggests a Gibbs sweep through the latent database $x$ will not suffer from poor mixing. 

\begin{algorithm}[htbp]
\caption{One iteration of the privacy-aware Metropolis-within-Gibbs sampler}
\label{algo:ims}
\begin{enumerate}
    \item\label{step:theta} Conditional update of $p(\theta \mid x)$ using the kernel from \Cref{assm:nonprivate}.
    \item\label{step:x}  For each $i = 1,2,\ldots$, sequentially update $x_i \mid x_{-i}, \theta, \sdp$.
        \begin{enumerate}
        \item\label{step:prop} Propose $x_i^{\star}\sim f(\cdot \mid \theta)$. 
        \item\label{step:update}  Update $t(x^{\star},s_{dp})=t(x,s_{dp})-t_i(x_i,s_{dp})+t_i(x_i^{\star},s_{dp})$ according to \Cref{assm:rec_add}. 
        \item\label{step:acc} Accept the proposed state with probability $\alpha(x_i^{\star} \mid x_i, x_{-i}, \theta)$ given by:
            \begin{equation}\label{eq:acceptance-prob}
   \hspace{-.15in}     \alpha(x_i^{\star} \mid x_i, x_{-i}, \theta) = \min\left\{\frac{\eta(s_{dp}\mid x_i^{\star}, x_{-i})}{\eta(s_{dp}\mid x_i, x_{-i})},1\right\} = \min\left\{\frac{g(\sdp, t(x^{\star},\sdp))}{g(\sdp, t(x,\sdp))},1\right\}.
    \end{equation}
    \end{enumerate}
\end{enumerate}
\end{algorithm}
%

Q2 recognizes that the conditionals $p(x_i|\theta,\sdp,x_{-i})$ are model- and mechanism-specific, and simulating from these is challenging in most settings. 
For this, we take the following simple approach in \Cref{algo:ims}: at each step, we propose $x_i$ from the model $f(x|\theta)$, 
and accept it with the appropriate Metropolis-Hastings
acceptance probability~\eqref{eq:acceptance-prob}.
%
Observe that our choice of proposal distribution is independent of the privacy mechanism, and 
ignores the privatized data $\sdp$ as well as all other elements $x_{-i}$.
Despite being simple and general-purpose, we show in~\Cref{prop:acceptance} that for $\epsilon-$DP, we can lower-bound the acceptance probability of proposals produced this way by $\exp(-\epsilon)$. 
%
%
%
This lower bound is key to efficiency:
despite the unconstrained nature of the proposal distribution, we can guarantee a minimum acceptance probability.
These two facts suggest our sampler will explore the space of databases relatively quickly.
\begin{restatable}{prop}{propacceptance}\label{prop:acceptance}
For a pure $\ep$-DP privacy mechanism $\eta$, 
the acceptance probability $\alpha$ from Equation \eqref{eq:acceptance-prob} satisfies
    $\alpha(x^{\star}_i \mid x_i, x_{-i}, \theta) \ge  \exp(-\ep),$ for all $\theta,x_{-i},x_i,x^*_i$.
\end{restatable}

The privacy loss budget $\ep$ is usually understood to be a small constant, which privacy experts recommend be between $.01$ and $1$ \citep{dwork2011firm}. When $\ep=1$, 
\Cref{prop:acceptance} ensures that the acceptance rate in \Cref{algo:ims} is no less than $36.7\%$, and as $\epsilon$ approaches zero, the bound on the acceptance rate approaches one. Intuitively, this is because as $\ep$ decreases, the distribution of the privatized data $\sdp$ depends less and less on any individual element of the database.



The simplicity of our approach arises through a decoupling of the data model from the privacy mechanism: the former is used to update $\theta$ and propose $x_i$'s, while the latter is used to calculate the acceptance probabilities.
The next result formalizes the computational efficiency of our approach. Specifically, for any record-additive mechanism, 
one iteration of our algorithm requires $O(n)$ operations, where $n$ is the size of the latent database. Essentially, this arises because of~\Cref{assm:rec_add}, which allows the acceptance probability in~\eqref{eq:acceptance-prob} to be calculated in $O(1)$ time. 
\begin{restatable}{prop}{propruntime}\label{prop:runtime}
The Gibbs sampler described in \Cref{algo:ims}
requires $O(n)$ number of operations to update the full latent database according to $p(x \mid \theta, \sdp)$.
\end{restatable}

Note that even without privacy, one round of an MCMC procedure typically takes $O(n)$ time. 
This is because updating $\theta$ given the confidential data requires computing the data likelihood $f(x\mid \theta) = \prod_{i=1}^n f(x_i \mid \theta)$, an $O(n)$ operation in general. 
Thus, as a result of the mild and typical condition that $\eta$ is record-additive, our MCMC procedure enjoys the \emph{same order} of runtime as the original MCMC algorithm for confidential data.

The previous two results are key to understanding the  efficiency of our approach. In the next section we formally establish geometric ergodicity of the sampler in \Cref{thm:ergodic_ims} and \ref{thm:geom_ergodic}.

{\bf Computational complexity.} 
The i.i.d.\ assumption on the records ensures that step~\ref{step:prop}  of Algorithm \ref{algo:ims} take $O(1)$ time, though this assumption can easily be weakened.
Assumption \ref{assm:rec_add} allows steps~\ref{step:update} and~\ref{step:acc} to also takes $O(1)$ time. 
Overall, step~\ref{step:x} of our algorithm then takes $O(n)$ (rather than $O(n^2)$) time as stated in Proposition \ref{prop:runtime}.
This matches the typical per-iteration cost of samplers for the non-private posterior distribution required in step~\ref{step:theta}. 
Thus, the overall cost of an iteration of our MCMC sampler is $O(n)$, which is typical when dealing with datasets of size $n$. 

\subsection{Ergodicity of the Privacy-Aware Sampler}\label{sec:ergodicity}

Ergodicity ensures the MCMC chain converges to the posterior distribution in total variation distance. 
\citep{tierney1994markov}
, which is essential for an MCMC sampler to 
 consistently estimate functionals of the posterior distribution. 
In \Cref{thm:ergodic_ims}, we provide mild and sufficient conditions for our proposed Metropolis-within-Gibbs
sampler to be ergodic. 

\begin{thm}\label{thm:ergodic_ims}
Under conditions A1 - A3 below, the Metropolis-within-Gibbs sampler of \Cref{algo:ims} 
on the joint space $(\mathbb{X}^{n} \times \Theta)$ is ergodic and it admits $p(x,\theta\mid \sdp)$ as the unique limiting distribution. 

 A1. The prior distribution is proper and $p(\theta) >0$ for all $\theta$ in $\Theta = \{\theta \mid f_{\theta}(x) > 0 \textnormal{ for some } x\}.$ \\
     A2. The model is such that the set $\{x: f(x\mid\theta)>0\}$ does not depend on $\theta.$\\
     A3. The privacy mechanism satisfies $\eta(\sdp \mid x) > 0 $ for  all $x \in \mathbb{X}^n.$ 
\end{thm}

We prove this in the supplementary material by verifying invariance, aperiodicity, and irreducibility. 
Conditions A1-A3 concern model specification, prior specification, and privacy noise. 
These mild assumptions are typically true and are easy to verify. 
While there are some mechanisms, such as the release-one-at-random mechanism, which satisfy approximate-DP but which violate A3 \citep{barber2014privacy}, most privacy mechanisms of interest satisfy A3. 
It is easy to verify that if $\eta$ satisfies $\ep$-DP, zero-concentrated DP, or Gaussian-DP, then property A3 is guaranteed. 

Next, we establish conditions for \Cref{algo:ims} 
to be geometrically ergodic~\citep{rosenthal1995minorization,roberts1998two}.
A chain is said to be \emph{geometrically ergodic} if its total variation distance to the 
target has a geometrically decaying upper bound. 
Geometric ergodicity is a desirable property since it provides a rate on convergence to the stationary distribution, guaranteeing central limit theorems, and 
allowing for the computation of asymptotically valid standard errors.

For simplicity, we focus on the situation where one can directly sample from the conditional posterior $p(\theta \mid x)$. 
This is an important and common case, relevant when either $\theta$ is low-dimensional or where one can place conjugate priors on $\theta$.
Both applications we present in this work,  a log-linear model in \Cref{sec:loglinear}, and a linear regression model in \Cref{sec:regression}, fall under this setting.

\begin{restatable}{thm}{geomergodic}\label{thm:geom_ergodic}
Assume that in step~\ref{step:theta} of~\Cref{algo:ims}, one can directly sample from 
$p(\theta \mid x)$.
Under A1-A3 of \Cref{thm:ergodic_ims}, the resulting $(x,\theta)$ chain, as well as the marginal chains, 
are geometrically ergodic if $\eta$ satisfies $\ep$-DP and 
there exists $0 < a \le b <\infty$ such that $a \le f(x \mid \theta) \le b \quad  \forall \theta, x.$
\end{restatable}

To prove Theorem \ref{thm:geom_ergodic}, we verify the drift and minorization conditions for component-wise Gibbs samplers 
in Theorem 8 of~\citep{johnson2013component}. See the supplementary material for details.  

Unsurprisingly, geometric ergodicity requires stronger assumptions than just ergodicity. 
The first assumption concerning the ability to sample directly from $p(\theta|x)$ can be avoided, but for the sake of clarity  we do not try to relax it, since we are mostly concerned with the interface with the privacy mechanism. 
The second assumption on the boundedness of the likelihood is stronger, but also typical. 
A common way to achieve bounded likelihoods is to require the sample space $\mathbb{X}$ (and typically also $\Theta$) to be bounded. 
In many real-world settings, such bounds exist, even if they may be very loose. 



\section{Na\"ive Bayes Log-Linear Model}\label{sec:loglinear}
Log-linear models are often used to model categorical data, 
a popular instance being the na\"ive Bayes classifier. 
Following \citet{Karwa2015:PrivatePD}, we consider the following model:
$x=(x_1,\ldots, x_K)$ is the input \emph{feature-vector}, with each $x_k$ taking values in $\{1,2,\ldots,J_k\}$, and $y$ is the output {\em class} taking values in $\{1,2,\ldots, I\}$.
Each input-output pair $(x,y)$ forms one record in our confidential database, the entire database consisting of $N$ i.i.d. copies of $(x,y)$. 
The na\"ive Bayes classifier assumes that $P(x\mid y) = \prod_{k=1}^K P(x_k\mid y)$, the model parameters being $p_{ij}^k=P(x_k=j|y=i)$ and $p_i=P(y=i)$. 
The sufficient statistics of this model are $n_{ij}^k = \#(y=i,x_k=j)$, which count the number of class-feature co-occurrences. 
%
This will form our confidential query $s$, which we privatize by adding Laplace noise to each of the $n_{ij}^k$. The resulting quantity $s_{dp}$, consisting of the noisy counts $m_{ij}^k=n_{ij}^k+L_{ijk},$ is what we release.
When $L_{ijk} \iid \mathrm{Laplace}(0,2K/\epsilon)$, the output $\sdp$ satisfies $\epsilon$-DP. 
%
Placing a $\mathrm{Dirichlet}(2,\ldots, 2)$ on all parameter vectors, our goal is to obtain the marginal posterior distribution of  $p,p_{i-}^k\mid s_{dp}$. 
While \citet{Karwa2015:PrivatePD} approximate this private posterior distribution using variational Bayes methods, our MCMC procedure is able to target the exact private posterior distribution.

{\bf Simulation setup.}
We perform several simulation experiments where we apply our MCMC samplers to the log-linear model described above. For the simulation, we set $N=100$ (number of records), 
$I=5$ (number of classes), $K=5$ (number of features), and $J_k=3$ for all $k=1,\ldots, K$ (possible values for each feature). 
We evaluate our sampler for privacy levels corresponding to $\epsilon \in \{.1,.3,1,3,10\}$.

\begin{figure}[htbp]
\centering
\begin{subfigure}[t]{.48\textwidth}
    \centering
    \includegraphics[width=\linewidth]{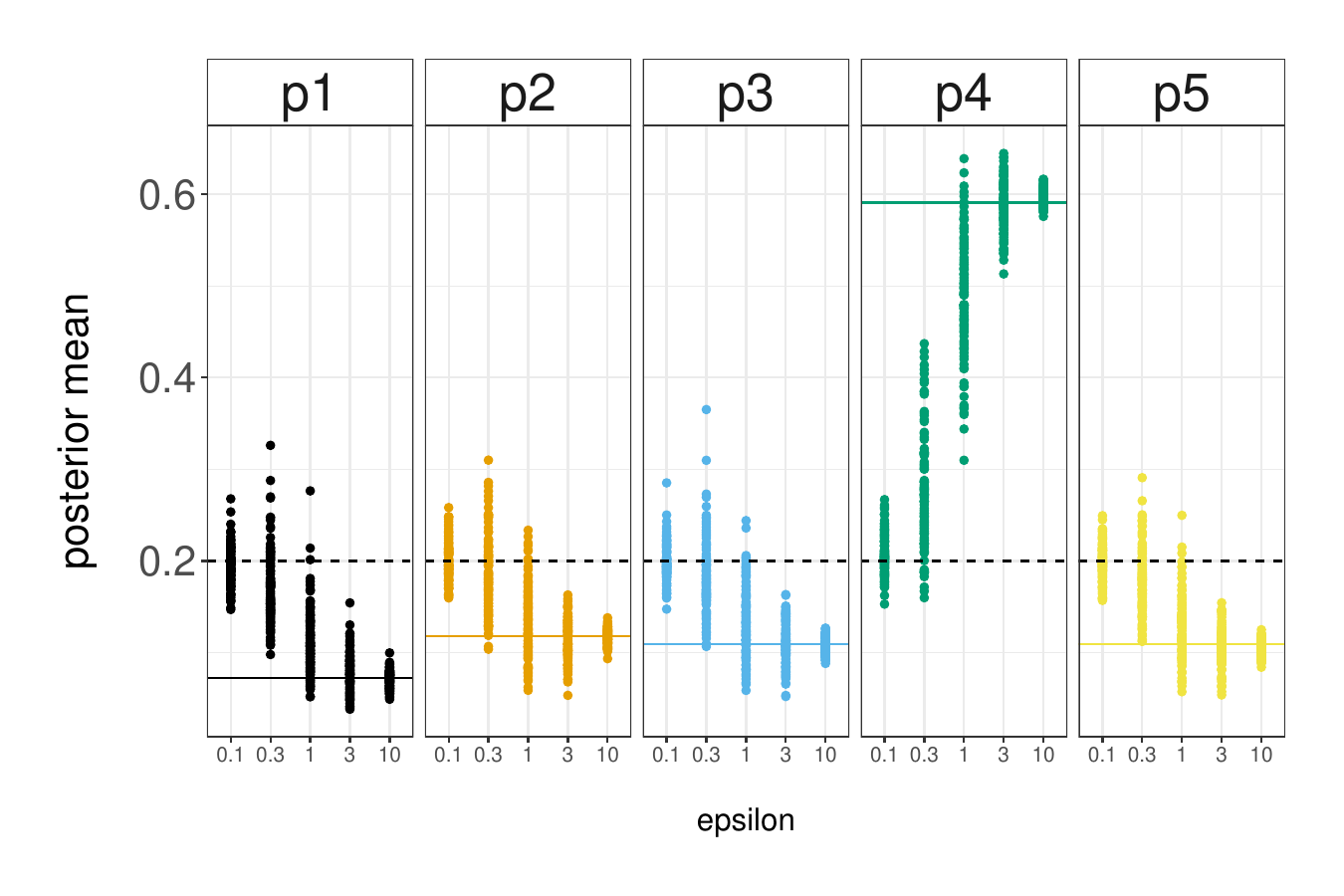}
    \caption{Posterior means for the log-linear model. The solid horizontal lines indicate the non-private posterior means, and the dashed lines at .2 indicate the prior means. }
    \label{fig:llpostmean.pdf}
\end{subfigure}
\hfill
\begin{subfigure}[t]{.48\textwidth}
    \centering
    \includegraphics[width=\linewidth]{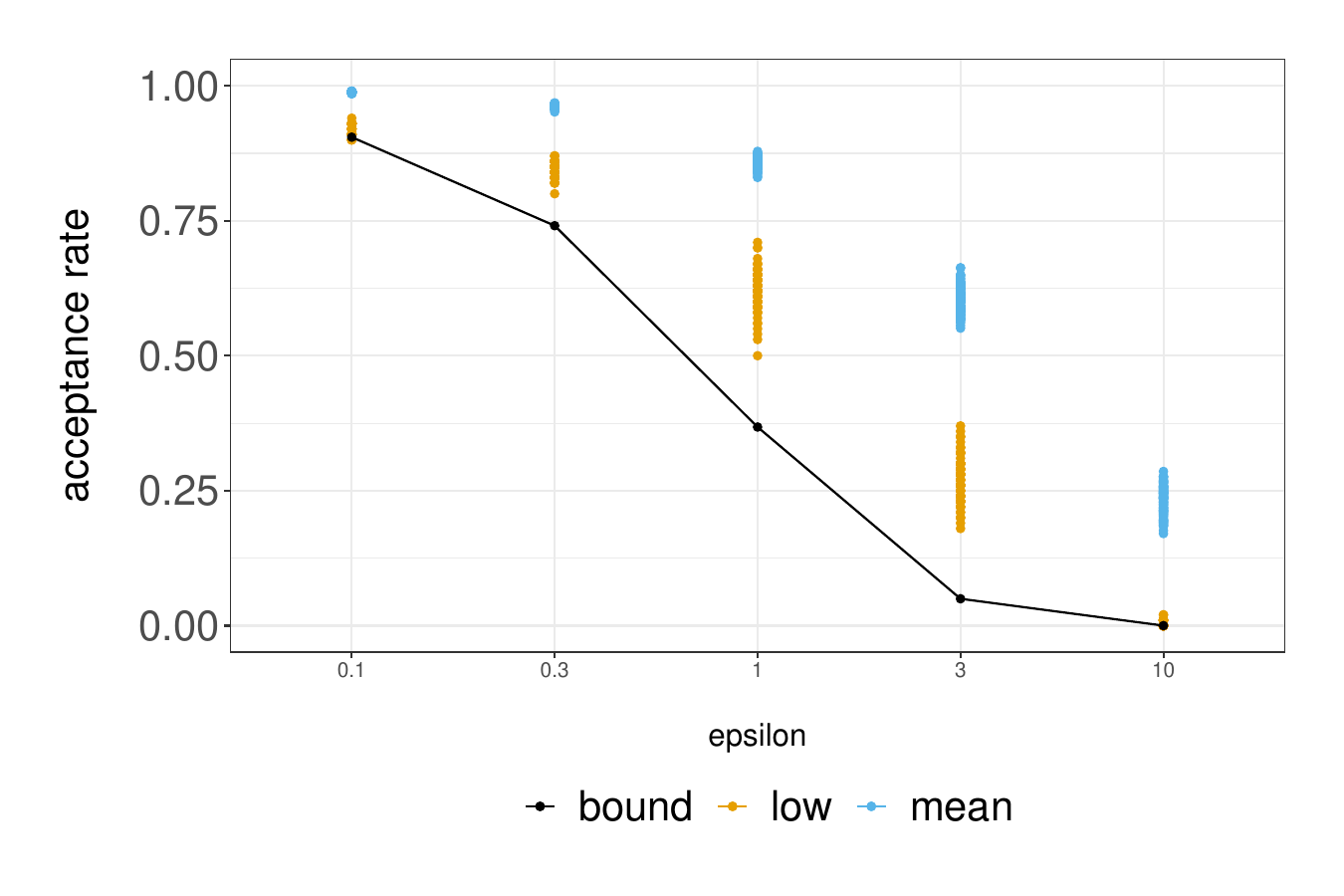}
    \caption{Observed acceptance rates for the log-linear model. The blue (above) point clouds indicate the average acceptance rate, and the orange (below) points indicate the observed minimum acceptance of each chain. The solid black line is the lower bound of~\Cref{prop:acceptance}.}
    \label{fig:llacceptance.pdf}
\end{subfigure}
\end{figure}

{\bf Posterior mean.} We generate one non-private dataset from the model, and hold it fixed. We then create 100 private queries $\sdp$ at each $\ep$ value, and for each $\sdp$ we run \Cref{algo:ims} for 10000 iterations. We discard the first 5000 iterations as burn-in. 
Finally, for each chain, we calculate the posterior mean. 
Figure \ref{fig:llpostmean.pdf} plots the 100 different posterior means for each $\ep$-value for the parameters $p_i = P(Y=i)$ for $i=1,\ldots, 5$. In this plot, the solid horizontal lines indicate the non-private posterior means, and the dashed horizontal lines indicate the prior means for each parameter. We see that as $\ep$ approaches zero, the posterior mean approaches the prior mean, reflecting the intuition that we learn less from the data as the privacy budget gets smaller. On the other hand, as $\ep$ increases, we see that the private posterior mean approaches the non-private posterior mean, which reflects the fact that as $\ep$ grows, we learn approximately the same from the data as if there were no privacy mechanism. 

{\bf Acceptance rate. } Using the same simulation setup as for the posterior mean, we calculate the average and minimum acceptance rate of Step~\ref{step:x} in \Cref{algo:ims}. 
Since the privacy mechanism satisfies $\ep$-DP, we know that $\exp(-\ep)$ is a lower bound on these acceptance rates. 
In Figure \ref{fig:llacceptance.pdf}, we confirm this bound, and see that the average acceptance rate is significantly higher than this lower bound. This suggests that the chain mixes even faster than indicated by \Cref{prop:acceptance}. 

{\bf Coverage of credible intervals. } For the next experiment, we sample a set of parameters from the prior, and hold this fixed. Then for each $\ep$ value, we produce 100 non-private datasets $(n_{ij}^k)$, one private dataset $(m_{ij}^k)$ for each non-private one, and then run a chain for 10,000 iterations discarding the first 5000 iterations for burn-in. From each chain, we produce a $90\%$ credible interval for each $p_i=P(Y=i)$, and calculate the empirical coverage which is reported in Table \ref{tab:llcoverage}. 
\begin{table}[htbp]
    \centering
      \[\begin{array}{c|ccccc}
            \ep & p_1=.097 & p_2=.148 & p_3=.145 & p_4=.446 & p_5 = .163 \\\hline
            .1&1&1&1&{\bf.36}&1\\
            .3&.97&1&1&{\bf .59}&1\\
            1&.94&.99&.97&{\bf.83}&.98\\
            3&.95&.91&.97&.89&.93\\
            10&.92&.88&.94&.92&.9
        \end{array}\]
        \caption{Coverage of $p_i=P(Y=i)$ for the log-linear model at different $\ep$. Top row is the true data generating parameter values. Coverage is based on 100 replicates.}
    \label{tab:llcoverage}
\end{table}

At a sample size of only $N=100$, we do not expect the coverage of the credible intervals to  match the nominal level of $90\%$, but we see in Table \ref{tab:llcoverage} that most of the coverage values are above $.9$.  Notable exceptions are the coverage of $p_4$ when $\ep$ is small. This may be because when $\ep$ is small, the private posterior is approximately equal to the prior, which is centered at $.2$; however $p_4$ is significantly further from $.2$ than the other parameters, which may explain why the coverage is low in this case. 

\section{Linear Regression}\label{sec:regression}
Next, we consider ordinary linear regression with $n$ subjects and $p$ predictors. 
We write $x_0$ for the matrix of predictors excluding the intercept columns, 
 $x = (\ul 1,x_0)$ for the matrix including the intercept, and 
$y$ for the vector of outcomes. 
We model the explanatory variables $x_0$ as
$x_0^i \iid \mathcal{N}_p(m,\Sigma)$ for $i = 1,\ldots,n =100$, with $y|x$ given by 
$\mathcal{N}_n(x\beta, \sigma^2 I_n)$.
Here $I_n$ is the $n\times n$ identity matrix and $\mathcal{N}_n$ denotes the $n$-dimentional multivariate Normal distribution.
The parameters of interest are $\beta$, the $(p+1)-$dimension vector of regression coefficients,
with $\sigma, m$ and $\Sigma$ assumed known.
We use independent $\mathcal{N}(0, 2^2)$ priors for the components of $\beta$.

To achieve $\ep$-DP via the Laplace mechanism, we require a finite global sensitivity. To achieve this, standard practice in the DP literature is to bound each predictor and response variable in a data-independent fashion. The bounds chosen by the privacy expert 
are $[a_i,b_i]$ for each instance of $x_0^i$ and $[a_y,b_y]$ for the entries of $y$, and these values are shared with the analyst. 

\begin{definition}
For a real value $z$, and $a\leq b$,  define the \emph{clamp function} $[z]_a^b \defeq \min\{\max\{z,a\},b\}$. If $z$ is a vector of length $d$, we use the same notation to apply an entry-wise clamp: $[z]_a^b \defeq ([z_1]_a^b,[z_2]_a^b,\ldots, [z_d]_a^b)^\top$. 
\end{definition}

Before adding noise for privacy, we first clamp the predictors and response, and then normalize them to take values in $[-1,1]$:
$  \twid x_0^i \defeq (b_i-a_i)^{-1}2([x_0^i]_{a_i}^{b_i}-a_i)-1$ and $\twid  y \defeq (b_y-a_y)^{-1}2([y]_{a_y}^{b_y}-a_y)-1.$
Call $\twid x \defeq [\ul 1,\twid x_0^1,\twid x_0^2,\ldots, \twid x_0^p]$ and $s\defeq(\twid x^\top \twid y, \twid y^\top \twid y, \twid x^\top \twid x)$. 
The $s$ is the summary statistic to which we will add noise for privacy. 
The $\ell_1$ sensitivity of $s$ (ignoring duplicate entries of $\twid x^\top \twid x$, and the constant entry $(\twid x^\top \twid x)_{1,1}$) is $\Delta = p^2+3p+3$. To satisfy $\ep$-DP, we add independent $\mathrm{Laplace}(0,\Delta/\ep)$ noise to each of the $d=\frac 12(p+1)(p+2)+(p+1)$ unique entries of $x$, which gives our final private summary $s_{dp}$.
We notice that $s$ is an additive function and each individual's contribution to $s$ is 
$t(x_i,y_i) = ( (\tilde{x}^i)^{\top} \tilde{y}_i, \tilde{y}_i^2, (\tilde{x}_i)^{\top} \tilde{x}_i).$
This mechanism producing $\sdp$ is record-additive.

{\bf Simulation setup.}
Our experiments focus on posterior inference about $\beta$ based on $S_{dp}$. 
For simplicity, we fix other parameters $\sas,m,\Sigma$ at the true data generating parameters (reported in the supplementary materials). 
When they are unknown, the posterior distributions of these parameters can be estimated by our Gibbs sampler as well.
Confidential predictors and responses are clamped with bounds $b = 10$ and $a = -10$. 
Given a confidential database $(x,y)$, the posterior distribution of $\beta$
is multivariate Normal and can be sampled directly with a runtime linear in $n$.

\begin{figure}[htbp]
\centering
    \includegraphics[width=.85\textwidth]{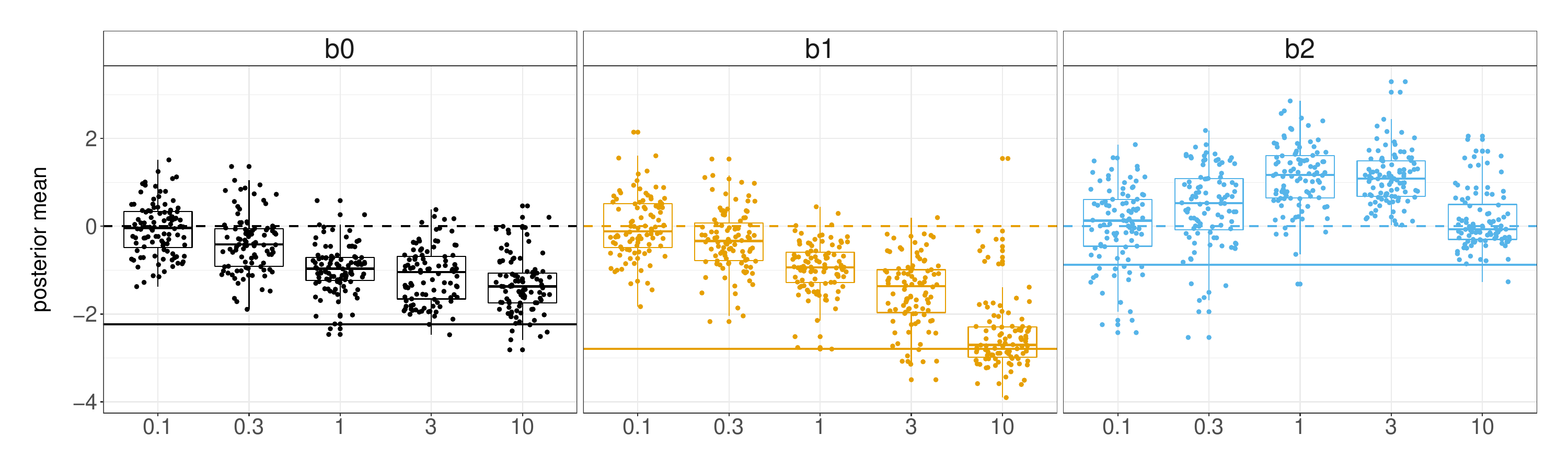}
    \caption{Posterior mean for private linear regression $\beta \mid \sdp$ with fixed confidential data. 
    The solid horizonal lines indicate the confidential data posterior means, and the dashed lines indicate 0.
    }
    \label{fig:lrpostmean}
\end{figure}

{\bf Posterior mean.} 
We generate one confidential dataset $(x,y)$ and hold it fixed. 
At each $\ep$ value, we create 100 private outputs $\sdp$ and run Gibbs samplers for 10,000 iterations targeting the 
posterior $\beta \mid \sdp$, discarding the first 5000 iterations.  
We plot the 100 different posterior means of $\beta$ in \Cref{fig:lrpostmean}.  
In this plot, the solid horizontal lines indicate posterior means given confidential data $(x,y)$, which we do not expect to fully recover due to clamping. 
The posterior quantities display the same trend with respect to change in privacy level as observed in \Cref{fig:llacceptance.pdf}. 
The other experiments from Section \ref{sec:loglinear} were also run on this linear regression model, and produced similar results. Simulation details, plots and discussion are in the supplementary materials.

\section{Discussion}\label{sec:discussion}
We proposed a novel, but simple sampling procedure for parameter inference  models where one only has access to privatized data. Our approach is an MCMC sampler that targets the posterior distribution $p(\theta \mid \sdp)$, and which leverages existing samplers for the non-private posterior  $p(\theta \mid x)$, as well as the structure of the privacy mechanism. The result is a simple wrapper for practitioners to obtain valid statistical inference from privatized data using the same models for the unobserved confidential data. As a side product, our algorithm also produces multiple copies of the 
confidential database from the posterior $p(x \mid \sdp)$. These posterior predictive draws could be useful when one is interested in inferring properties of $x$ as well.
Although we did not discuss this, our data augmentation scheme can also potentially enable frequentist analysis through the Monte Carlo expectation-maximization algorithm. 
  
We acknowledge some limitations of present work. First, we point out that strong assumptions such as bounded parameter space $\Theta$ and sample space $\mathbb{X}$ are required to establish geometric ergodicity of the Gibbs sampler in \Cref{thm:geom_ergodic}.
They mostly reflect the current state of MCMC mixing results, and can likely be relaxed, at the cost of a more complex theorem statement and proof. 
Second, our current proposal for updating $x_i \mid x_{-i},\theta,\sdp$ only tailors to the model $f(\cdot \mid \theta)$ and it is not customized for $\sdp$ yet. In the future, we might be able to design algorithms that also incorporate the privatized output $\sdp$ in these proposals.
Third, we point out while our method exploits privacy to control the correlation between the components of $x$, it is still susceptible to correlation between $x$ and $\theta$. This  can potentially cause poor mixing in practice, despite geometric convergence rate. 
While in simple problems, this can be fixed by reparameterization, we plan to develop MCMC algorithms for  this setting in
follow-up studies. 
A similar potential issue is that, as a wrapper, our algorithm may be inefficient when the chosen sampler for $p(\theta \mid x)$ does not mix well, especially in some high-dimensional settings. We emphasize though that our method is intended for settings where there currently exists efficient samplers for $p(\theta\mid x)$, and when this is not the case, alternative approaches may be needed. 

Finally, while the proposed algorithm converges so long as the privacy mechanism $\eta$ is known,
irrespective of the specific privacy guarantee,
we point out that for alternative versions of DP (such as zCDP), the acceptance probability results in \Cref{prop:acceptance} may no longer hold, as this result depends on the $\ep$-DP guarantee. Developing alternatives to \Cref{prop:acceptance} for different privacy definitions is another goal of future work.



\begin{ack}

R.\ Gong is supported in part by the National Science Foundation grant DMS-1916002,  V.\ Rao by the National Science Foundation grants RI-1816499 and DMS-1812197, and J.\ Awan by the National Science Foundation grant SES-2150615.


\end{ack}

\bibliographystyle{plainnat}
\bibliography{{arxiv_v1}}

\begin{thebibliography}{36}
\providecommand{\natexlab}[1]{#1}
\providecommand{\url}[1]{\texttt{#1}}
\expandafter\ifx\csname urlstyle\endcsname\relax
  \providecommand{\doi}[1]{doi: #1}\else
  \providecommand{\doi}{doi: \begingroup \urlstyle{rm}\Url}\fi

\bibitem[Awan and Slavkovi{\'c}(2018)]{Awan2018:Binomial}
Jordan~Alexander Awan and Aleksandra Slavkovi{\'c}.
\newblock Differentially private uniformly most powerful tests for binomial
  data.
\newblock In S.~Bengio, H.~Wallach, H.~Larochelle, K.~Grauman, N.~Cesa-Bianchi,
  and R.~Garnett, editors, \emph{Advances in Neural Information Processing
  Systems 31}, pages 4208--4218, 2018.

\bibitem[Awan and Slavkovi{\'c}(2020)]{awan2020differentially}
Jordan~Alexander Awan and Aleksandra Slavkovi{\'c}.
\newblock Differentially private inference for binomial data.
\newblock \emph{Journal of Privacy and Confidentiality}, 10\penalty0 (1), 2020.

\bibitem[Barber and Duchi(2014)]{barber2014privacy}
Rina~Foygel Barber and John~C Duchi.
\newblock Privacy and statistical risk: Formalisms and minimax bounds.
\newblock \emph{arXiv preprint arXiv:1412.4451}, 2014.

\bibitem[Bernstein and Sheldon(2018)]{bernstein2018differentially}
Garrett Bernstein and Daniel~R Sheldon.
\newblock Differentially private bayesian inference for exponential families.
\newblock \emph{Advances in Neural Information Processing Systems},
  31:\penalty0 2919--2929, 2018.

\bibitem[Bernstein and Sheldon(2019)]{bernstein2019differentially}
Garrett Bernstein and Daniel~R Sheldon.
\newblock Differentially private bayesian linear regression.
\newblock \emph{Advances in Neural Information Processing Systems},
  32:\penalty0 525--535, 2019.

\bibitem[Bun and Steinke(2016)]{bun2016concentrated}
Mark Bun and Thomas Steinke.
\newblock Concentrated differential privacy: Simplifications, extensions, and
  lower bounds.
\newblock In \emph{Theory of Cryptography Conference}, pages 635--658.
  Springer, 2016.

\bibitem[Cai et~al.(2021)Cai, Wang, and Zhang]{cai2019cost}
T~Tony Cai, Yichen Wang, and Linjun Zhang.
\newblock The cost of privacy: Optimal rates of convergence for parameter
  estimation with differential privacy.
\newblock \emph{The Annals of Statistics}, 49\penalty0 (5):\penalty0
  2825--2850, 2021.

\bibitem[Chaudhuri et~al.(2011)Chaudhuri, Monteleoni, and
  Sarwate]{Chaudhuri2011}
Kamalika Chaudhuri, Claire Monteleoni, and D.~Sarwate.
\newblock Differentially private empirical risk minimization.
\newblock In \emph{Journal of Machine Learning Research}, volume~12, pages
  1069--1109, 2011.

\bibitem[Dalenius(1977)]{Dalenius1977}
Tore Dalenius.
\newblock Towards a methodology for statistical disclosure control.
\newblock \emph{Statistik Tidskrift}, 15:\penalty0 429--444, 1977.

\bibitem[Dong et~al.(2021)Dong, Roth, and Su]{dong2021gaussian}
Jinshuo Dong, Aaron Roth, and Weijie Su.
\newblock Gaussian differential privacy.
\newblock \emph{Journal of the Royal Statistical Society, Series B}, 2021.

\bibitem[Dwork(2011)]{dwork2011firm}
Cynthia Dwork.
\newblock A firm foundation for private data analysis.
\newblock \emph{Communications of the ACM}, 54\penalty0 (1):\penalty0 86--95,
  2011.

\bibitem[Dwork and Rothblum(2016)]{dwork2016concentrated}
Cynthia Dwork and Guy~N Rothblum.
\newblock Concentrated differential privacy.
\newblock \emph{arXiv preprint arXiv:1603.01887}, 2016.

\bibitem[Dwork et~al.(2006)Dwork, McSherry, Nissim, and
  Smith]{dwork2006calibrating}
Cynthia Dwork, Frank McSherry, Kobbi Nissim, and Adam Smith.
\newblock Calibrating noise to sensitivity in private data analysis.
\newblock In \emph{Theory of cryptography conference}, pages 265--284.
  Springer, 2006.

\bibitem[Ferrando et~al.(2022)Ferrando, Wang, and
  Sheldon]{ferrando2022parametric}
Cecilia Ferrando, Shufan Wang, and Daniel Sheldon.
\newblock Parametric bootstrap for differentially private confidence intervals.
\newblock In \emph{International Conference on Artificial Intelligence and
  Statistics}, pages 1598--1618. PMLR, 2022.

\bibitem[Gaboardi and Rogers(2018)]{gaboardi2018local}
Marco Gaboardi and Ryan Rogers.
\newblock Local private hypothesis testing: Chi-square tests.
\newblock In \emph{International Conference on Machine Learning}, pages
  1626--1635. PMLR, 2018.

\bibitem[Gaboardi et~al.(2016)Gaboardi, Lim, Rogers, and
  Vadhan]{gaboardi2016differentially}
Marco Gaboardi, Hyun Lim, Ryan Rogers, and Salil Vadhan.
\newblock Differentially private chi-squared hypothesis testing: Goodness of
  fit and independence testing.
\newblock In \emph{International conference on machine learning}, pages
  2111--2120. PMLR, 2016.

\bibitem[Gilks et~al.(1995)Gilks, Best, and Tan]{gilks1995adaptive}
Wally~R Gilks, Nicky~G Best, and Keith~KC Tan.
\newblock Adaptive rejection {Metropolis} sampling within {Gibbs} sampling.
\newblock \emph{Journal of the Royal Statistical Society: Series C (Applied
  Statistics)}, 44\penalty0 (4):\penalty0 455--472, 1995.

\bibitem[Gong(2022)]{gong2019exact}
Ruobin Gong.
\newblock Exact inference with approximate computation for differentially
  private data via perturbations.
\newblock \emph{Journal of Privacy and Confidentiality (to appear)}, 2022.

\bibitem[Johnson et~al.(2013)Johnson, Jones, and Neath]{johnson2013component}
Alicia~A Johnson, Galin~L Jones, and Ronald~C Neath.
\newblock Component-wise {Markov} chain {Monte} {Carlo}: {Uniform} and
  geometric ergodicity under mixing and composition.
\newblock \emph{Statistical Science}, 28\penalty0 (3):\penalty0 360--375, 2013.

\bibitem[Karwa et~al.(2016)Karwa, Kifer, and Slavkovi\'c]{Karwa2015:PrivatePD}
Vishesh Karwa, Dan Kifer, and Aleksandra Slavkovi\'c.
\newblock Private posterior distributions from variational approximations.
\newblock \emph{NIPS 2015 Workshop on Learning and Privacy with Incomplete Data
  and Weak Supervision}, 2016.

\bibitem[Karwa et~al.(2017)Karwa, Krivitsky, and
  Slavkovi{\'c}]{karwa2017sharing}
Vishesh Karwa, Pavel~N Krivitsky, and Aleksandra~B Slavkovi{\'c}.
\newblock Sharing social network data: differentially private estimation of
  exponential family random-graph models.
\newblock \emph{Journal of the Royal Statistical Society: Series C (Applied
  Statistics)}, 66\penalty0 (3):\penalty0 481--500, 2017.

\bibitem[Kifer et~al.(2012)Kifer, Smith, and Thakurta]{Kifer2012:PrivateCERM}
Daniel Kifer, Adam Smith, and Abhradeep Thakurta.
\newblock Private convex empirical risk minimization and high-dimensional
  regression.
\newblock In Shie Mannor, Nathan Srebro, and Robert~C. Williamson, editors,
  \emph{Proceedings of the 25th Annual Conference on Learning Theory},
  volume~23 of \emph{Proceedings of Machine Learning Research}, Edinburgh,
  Scotland, 25--27 Jun 2012. PMLR.

\bibitem[McSherry and Talwar(2007)]{mcsherry2007mechanism}
Frank McSherry and Kunal Talwar.
\newblock Mechanism design via differential privacy.
\newblock In \emph{48th Annual IEEE Symposium on Foundations of Computer
  Science (FOCS'07)}, pages 94--103. IEEE, 2007.

\bibitem[Murray et~al.(2012)Murray, Ghahramani, and MacKay]{murray2012mcmc}
Iain Murray, Zoubin Ghahramani, and David MacKay.
\newblock {MCMC} for doubly-intractable distributions.
\newblock \emph{arXiv preprint arXiv:1206.6848}, 2012.

\bibitem[Reimherr and Awan(2019)]{reimherr2019kng}
Matthew Reimherr and Jordan Awan.
\newblock {KNG}: the k-norm gradient mechanism.
\newblock \emph{Advances in Neural Information Processing Systems}, 32, 2019.

\bibitem[Roberts and Rosenthal(1998)]{roberts1998two}
Gareth~O Roberts and Jeffrey~S Rosenthal.
\newblock Two convergence properties of hybrid samplers.
\newblock \emph{The Annals of Applied Probability}, 8\penalty0 (2):\penalty0
  397--407, 1998.

\bibitem[Rosenthal(1995)]{rosenthal1995minorization}
Jeffrey~S Rosenthal.
\newblock Minorization conditions and convergence rates for {Markov} chain
  {Monte} {Carlo}.
\newblock \emph{Journal of the American Statistical Association}, 90\penalty0
  (430):\penalty0 558--566, 1995.

\bibitem[Schein et~al.(2019)Schein, Wu, Schofield, Zhou, and
  Wallach]{pmlr-v97-schein19a}
Aaron Schein, Zhiwei~Steven Wu, Alexandra Schofield, Mingyuan Zhou, and Hanna
  Wallach.
\newblock Locally private {Bayesian} inference for count models.
\newblock In \emph{Proceedings of the 36th International Conference on Machine
  Learning}, Proceedings of Machine Learning Research, pages 5638--5648. PMLR,
  09--15 Jun 2019.

\bibitem[Smith(2011)]{smith2011privacy}
Adam Smith.
\newblock Privacy-preserving statistical estimation with optimal convergence
  rates.
\newblock In \emph{Proceedings of the forty-third annual ACM symposium on
  Theory of computing}, pages 813--822, 2011.

\bibitem[Tanner and Wong(1987)]{tanner1987calculation}
Martin~A Tanner and Wing~Hung Wong.
\newblock The calculation of posterior distributions by data augmentation.
\newblock \emph{Journal of the American statistical Association}, 82\penalty0
  (398):\penalty0 528--540, 1987.

\bibitem[Tierney(1994)]{tierney1994markov}
Luke Tierney.
\newblock {Markov} chains for exploring posterior distributions.
\newblock \emph{the Annals of Statistics}, pages 1701--1728, 1994.

\bibitem[Van~Dyk and Meng(2001)]{van2001art}
David~A Van~Dyk and Xiao-Li Meng.
\newblock The art of data augmentation.
\newblock \emph{Journal of Computational and Graphical Statistics}, 10\penalty0
  (1):\penalty0 1--50, 2001.

\bibitem[Wang et~al.(2015)Wang, Lee, and Kifer]{wang2015revisiting}
Yue Wang, Jaewoo Lee, and Daniel Kifer.
\newblock Revisiting differentially private hypothesis tests for categorical
  data.
\newblock \emph{arXiv preprint arXiv:1511.03376}, 2015.

\bibitem[Wang et~al.(2018)Wang, Kifer, Lee, and Karwa]{wang2018statistical}
Yue Wang, Daniel Kifer, Jaewoo Lee, and Vishesh Karwa.
\newblock Statistical approximating distributions under differential privacy.
\newblock \emph{Journal of Privacy and Confidentiality}, 8\penalty0 (1), 2018.

\bibitem[Williams and McSherry(2010)]{williams2010probabilistic}
Oliver Williams and Frank McSherry.
\newblock Probabilistic inference and differential privacy.
\newblock \emph{Advances in Neural Information Processing Systems},
  23:\penalty0 2451--2459, 2010.

\bibitem[Zhang et~al.(2012)Zhang, Zhang, Xiao, Yang, and Winslett]{Zhang2012}
Jun Zhang, Zhenjie Zhang, Xiaokui Xiao, Yin Yang, and Marianne Winslett.
\newblock Functional mechanism: Regression analysis under differential privacy.
\newblock \emph{Proc. VLDB Endow.}, 5\penalty0 (11):\penalty0 1364--1375, July
  2012.
\newblock ISSN 2150-8097.
\newblock \doi{10.14778/2350229.2350253}.

\end{thebibliography}

\newpage
\section*{Checklist}

\begin{enumerate}

\item For all authors...
\begin{enumerate}
  \item Do the main claims made in the abstract and introduction accurately reflect the paper's contributions and scope?
    \answerYes{}
  \item Did you describe the limitations of your work?
    \answerYes{See details in \Cref{sec:discussion}.}
  \item Did you discuss any potential negative societal impacts of your work?
    \answerNA{} We do not foresee direct negative societal impact from the current work. See discussions in \Cref{sec:societal_impact}. 
  \item Have you read the ethics review guidelines and ensured that your paper conforms to them?
    \answerYes{}
\end{enumerate}

\item If you are including theoretical results...
\begin{enumerate}
  \item Did you state the full set of assumptions of all theoretical results?
    \answerYes{}
        \item Did you include complete proofs of all theoretical results?
    \answerYes{See \Cref{sec:proof1,sec:proof2}}.
\end{enumerate}

\item If you ran experiments...
\begin{enumerate}
  \item Did you include the code, data, and instructions needed to reproduce the main experimental results (either in the supplemental material or as a URL)?
    \answerYes{We believe that we have provided sufficient details to replicate our experiments in relevant sections. We will release our code to a public GitHub repository prior to the conference.}
    \item Did you specify all the training details (e.g., data splits, hyperparameters, how they were chosen)?
    \answerYes{We described how the prior hyper-parameters are chosen in relevant sections. In short, we choose proper and conjugate prior distributions for the log-linear and linear regression models. We also avoid informative priors.}
    \item Did you report error bars (e.g., with respect to the random seed after running experiments multiple times)?
    \answerNo{We directly display all the data points in the figures.}
    \item Did you include the total amount of compute and the type of resources used (e.g., type of GPUs, internal cluster, or cloud provider)?
    \answerYes{See Appendix.}
\end{enumerate}

\item If you are using existing assets (e.g., code, data, models) or curating/releasing new assets...
\begin{enumerate}
  \item If your work uses existing assets, did you cite the creators?
    \answerNA{}
  \item Did you mention the license of the assets?
    \answerNA{}
  \item Did you include any new assets either in the supplemental material or as a URL?
    \answerNA{}
  \item Did you discuss whether and how consent was obtained from people whose data you're using/curating?
    \answerNA{}
  \item Did you discuss whether the data you are using/curating contains personally identifiable information or offensive content?
    \answerNA{}
\end{enumerate}

\item If you used crowdsourcing or conducted research with human subjects...
\begin{enumerate}
  \item Did you include the full text of instructions given to participants and screenshots, if applicable?
    \answerNA{}
  \item Did you describe any potential participant risks, with links to Institutional Review Board (IRB) approvals, if applicable?
    \answerNA{}
  \item Did you include the estimated hourly wage paid to participants and the total amount spent on participant compensation?
    \answerNA{}
\end{enumerate}
\end{enumerate}
\vfill

\pagebreak
\appendix
\begin{center}
\textbf{\large Supplemental Materials: Data Augmentation MCMC for\\
Bayesian Inference from Privatized Data}
\end{center}
\setcounter{equation}{0}
\setcounter{figure}{0}
\setcounter{table}{0}
\setcounter{page}{1}
\setcounter{section}{0}
\makeatletter
\renewcommand{\thesection}{S-\arabic{section}}
\renewcommand{\theequation}{S\arabic{equation}}
\renewcommand{\thefigure}{S\arabic{figure}}
\renewcommand{\bibnumfmt}[1]{[S#1]}
\renewcommand{\citenumfont}[1]{S#1}

\section{Statement on Societal Impacts}\label{sec:societal_impact}
We do not foresee direct negative societal impact from the current work. 
Admittedly, our method is based on imputing the confidential database which privacy mechanisms seek to protect. 
We can assure the reader that such imputations are based on formally differentially private data products and
hence do not violate differential privacy. 
Also, one may argue that our work is catalytic to enhancing the `disclosure risk' of individuals, i.e. an adversary might be able to make accurate posterior inference about an individual if the adversary has highly informative and correct prior and modeling information to begin with. Granted, no existing privacy frameworks can guard against this. 

\section{Proofs in \texorpdfstring{\Cref{sec:gibbs-overview}}{gibbs-overview}}\label{sec:proof1}

\propacceptance*

\begin{proof}
Step~\ref{step:prop} of ~\Cref{algo:ims} proposes a new state $x_i^{\star}$ for the i-th record $x_i$ according to the model $f(\cdot \mid \theta)$. 
Notice that the proposed latent database $x^{\star} = (x_i^{\star}, x_{-i})$ and the current latent database $x = (x_i,x_{-i})$ differ in only one entry.
Then, the probability of accepting a proposed state
$x_i^{\star}$ is 
$\alpha(x^{\star}_i \mid x_i, x_{-i}, \theta) = \min\left(\eta_{\epsilon}(s_{dp}\mid x^{\star}) /  \eta_{\epsilon}(s_{dp}\mid x) ,1\right)$. 
This ratio compares two adjacent databases $x^{\star}$ and $x$.
$\epsilon$-DP guarantees that the probability ratio of any output is within $\exp(\pm \ep)$ for adjacent databases
by~\Cref{eq:eta-ratio}.
\end{proof}

\propruntime*
\begin{proof}
We prove that each update for $x_i \mid x_{-i},\theta,\sdp$ is $O(1)$ and hence the full sweep for the latent database $x \mid \theta,\sdp$ is $O(n)$.
Given current state $(x,\theta)$, in Step~\ref{step:prop}, the method proposes from $x_i^{\star} \sim f(\cdot \mid \theta)$ independent of other entries $x_{-i}$ and the current state $x_i$; the runtime of this local proposal step does not depend on $n$. 
Since $\eta(\sdp \mid x)$ is record-additive (\Cref{assm:rec_add}), then $t(x^{\star}, \sdp)$ can be computed in $O(1)$ time by $t(x^{\star},\sdp) = t(x,\sdp) - t_i(x_i,\sdp) + t_i(x_i^{\star}, \sdp)$ of Step~\ref{step:update}. The density evaluations in Step~\ref{step:acc} are also $O(1)$. Overall, to update all $x_i,\ i=1,2,\ldots,n$, the runtime is $O(n).$
\end{proof}

\section{Proofs in \texorpdfstring{\Cref{sec:ergodicity}}{ergodicity}}\label{sec:proof2}

\subsection{Ergodicity}

In \Cref{algo:metropolis-within-gibbs}, we first present a Metropolis-within-Gibbs sampler that is more general than \Cref{algo:ims}. We prove its ergodicity in \Cref{thm:ergodic}, which implies \Cref{thm:ergodic_ims}.

\begin{algorithm}[htbp]
\caption{A general Metropolis-within-Gibbs sampler for $p(\theta, x \mid \sdp)$}\label{algo:metropolis-within-gibbs}
\begin{enumerate}
    \item Conditional update of $p(\theta \mid x)$:
    \begin{enumerate}
    \item Propose $\theta^{\star}\sim q_\theta(\theta^{\star}\mid \theta, x)$. 
    \item Accept $\theta^{\star}$ with probability
    \[\alpha(\theta^{\star} \mid \theta, x) = \min\left\{\frac{q_\theta(\theta\mid \theta^{\star},x) p(\theta^{\star}) \prod_{i=1}^n f(x_i \mid \theta^{\star})}{q_{\theta}(\theta^{\star}\mid \theta,x) p(\theta) \prod_{i=1}^n f(x_i \mid \theta) },1\right\}\]
    \end{enumerate}
    \item For each $i=1,\ldots, n$, update $p(x_i \mid x_{-i}, \theta, \sdp)$ by: 
    \begin{enumerate}
        \item Propose $x_i'\sim q_x(x_i^{\star}\mid x_i, x_{-i}, \theta, \sdp)$,
        \item Accept the proposed state $x_i^{\star}$ with probability 
        \[\min\left\{\frac{q_x(x_i\mid x_i^{\star},x_{-i}, \theta,\sdp) \eta(s_{dp} \mid x_i^{\star}, x_{-i})f(x_i^{\star}\mid \theta)}{q_x(x_i^{\star}\mid x_i,x_{-i}, \theta,\sdp) \eta(s_{dp} \mid x_i, x_{-i}) f(x_i\mid \theta)},1\right\}.\]
    \end{enumerate}
\end{enumerate}
\end{algorithm}

The Metropolis-within-Gibbs sampler in \Cref{algo:metropolis-within-gibbs} consists of alternating Metropolis-Hastings steps targeting $p(\theta \mid x,\sdp) = p(\theta \mid x)$ and $p(x \mid \theta, \sdp)$. 
In \Cref{assm:nonprivate} we have assumed that a Markov kernel for $p(\theta \mid x)$ exists. A typical kernel involves first proposing from some distribution $q_{\theta}(\theta \mid x)$ and then accepting or rejecting the proposed state an appropriate probability. The data-augmentation steps consist of the sequence of updates $p(x_i |  x_{-i}, \theta, \sdp)$, for $i = 1,2,\ldots, n$. 
\Cref{algo:ims} suggests using the proposal $x_i^{\star}\sim f(\cdot \mid \theta)$ independent of current state $x$. In this more general sampler, described in \cref{algo:metropolis-within-gibbs}, we use proposals $q_x(x_i^{\star} \mid x_i, x_{-i},\theta,\sdp)$ that can depend on current states of $x$ and $\theta$, as well as the private query $\sdp.$ Notice that since latent records are exchangeable in both $f(x\mid \theta)$ and $\eta(\sdp \mid x)$, respectively by the i.i.d. model assumption and by record-additivity, it is sufficient to use the same kernel $q_x$ for all $x_i$. 

\begin{restatable}{thm}{thmergodic}\label{thm:ergodic}
Under conditions A1 - A4 below, the Gibbs sampler of \Cref{algo:metropolis-within-gibbs}
on the joint space $(\mathbb{X}^{n} \times \mathbb{R}^p)$ is ergodic and it admits $\pi(x,\theta)$ as the unique limiting distribution. 
\begin{itemize}
    \item[A1.] The prior distribution is proper and $\pi_0(\theta) >0$ for all $\theta$ in $\Theta = \{\theta \mid f_{\theta}(x) > 0 \textnormal{ for some } x\}.$ 
    \item[A2.] The model is such that the set $\{x: f(x\mid\theta)>0\}$ does not depend on $\theta.$
    \item[A3.] The privacy mechanism satisfies $\eta(\sdp \mid x) > 0 $ for  all $x \in \mathbb{X}^n.$ 
    \item[A4.] From a valid current state, the proposal kernels satisfies (a) 
    $q_{\theta}(\theta^{\star} \mid x,\theta) > 0$ for all $\theta^{\star} \in \Theta$, and (b) $q_{x}(x_i^{\star} \mid x_i,x_{-i}, \theta, \sdp) > 0$ for all $x_i^{\star}$ with $f(x_i^{\star} , x_{-i} \mid \theta) > 0.$
\end{itemize} 
\end{restatable}

\begin{proof}
It is sufficient to show that the chain is $\pi$-invariant, aperiodic, and $\pi$-irreducible~\citep{tierney1994markov}.
The Metropolis-within-Gibbs sampler is aperiodic by construction, 
since some proposals can be rejected. 
It is also $\pi$-invariant because it is composed of kernels that satisfy detailed balance with respect to $\pi$. 

Irreducibility means that, informally, every set $A$ with $\pi(A) > 0$ 
can be reached by the Gibbs sampler from any starting point within finitely many steps. 
We first prove irreducibility for $n = 1$ and generalize this to a  sample size of  $n \ge 2.$
Suppose $A \subset \mathbb{X}^{1} \times \Theta$ 
with $\pi(A) > 0$ and suppose the current state of the Gibbs chain is $(x^{(0)}, \theta^{(0)})$. For any state $(x,\theta) \in A$ we have $q(\theta \mid x^{(0)},x^{(0)}) q(x \mid x^{(0)},\theta, \sdp) >0$ by A4. The acceptance ratios are also positive by A1-A4. As a result 
\begin{align*}
&P(A \mid x^{(0)}, \theta^{(0)}) \\
&\ge \int\int_{A}  q(\theta \mid x^{(0)},x^{(0)}) q(x \mid x^{(0)},\theta,\sdp) 
\alpha(\theta \mid x^{(0)},x^{(0)}) \alpha(x \mid x^{(0)},\theta,\sdp)
dx d\theta >0.
\end{align*}
So when $n = 1$, we can reach $A$ from any starting point in one iteration of the Gibbs sampler. 
For $n \ge 2$, we can reach the set $A$ in at most $n$ steps: the first iteration moves $x_1$ and $\theta$ into $A$, and subsequent steps moves other $x_i$'s into $A$ while keeping all previous $x_j$'s inside $A$ by rejecting proposals leaving $A$.
\end{proof}

A4 details conditions on the proposal distributions to ensure ergodicity of \Cref{algo:metropolis-within-gibbs}. It can be relaxed so long as $\pi$-irreducibility is satisfied. Also, A4a should be viewed as a condition implied by the validility of a kernel targeting $p(\theta \mid x)$ from \Cref{assm:nonprivate} and, therefore, is not  an additional assumption. 
Importantly, 
conditions in A4 are mild because they cover common proposal distributions; Gaussian random walk on $\theta$ for A4a and the independent Metropolis proposals $f(\cdot \mid \theta)$ for A4b are such examples. 
In \Cref{algo:ims}, we use the kernel $q_x(x_i^{\star} \mid x_i,x_{-i},\theta,\sdp) = f(x \mid \theta)$, which satisfies $f(x^{\star} \mid \theta)  > 0$ by A2. Hence \Cref{thm:ergodic} implies \Cref{thm:ergodic_ims}.

\subsection{Geometric ergodicity of \texorpdfstring{\Cref{algo:ims}}{algo:ims}}

\geomergodic*

\begin{proof}[Proof of \Cref{thm:geom_ergodic}]
The assumption of $a \le f(x \mid \theta) \le b$ leads to the inequality
\begin{equation*}
    p(\theta \mid x) = \frac{p(\theta) f(x \mid \theta)}{ \int p(\theta') f(x \mid \theta') d\theta'} \ge \frac{a}{b}p(\theta),
\end{equation*} since $p(\theta)$ is a proper prior by A1 of \cref{algo:ims}. 

This proof proceeds by verifying the drift and minorization conditions of the marginal Markov transition kernel on $X$ according to Theorem 8 of~\citet{johnson2013component}.
We first present a full proof for $n = 1$ and then generalize the arguments to $n \ge 2$. In this proof, we abbreviate $\eta(\sdp \mid x)$ as $\eta(x).$

Recall that the probability of accepting proposed state $x^{\star}$ is 
$\alpha(x^{\star} \mid x,\theta) = \min\left(1,\frac{\eta(x^{\star})}{\eta(x)}\right).$ The probability of accepting any proposal from the current state is $\alpha(x,\theta) = \int \alpha(x^{\star} \mid x,\theta) f(x^{\star} \mid \theta) dx^{\star}.$ %
Let $K(x' \mid x,\theta)$ denote the Markov transition kernel with respect to the proposal $x^{\star} \sim f(\cdot \mid \theta)$, and let $K(x' \mid x) = \int K(x' \mid x,\theta) p(\theta \mid x) d\theta$ be the marginal kernel,  which integrates out the exact $\theta$ update from $p(\theta \mid x).$
We have 
$$ K(x' \mid x,\theta) = f(x' \mid \theta)\alpha(x' \mid x,\theta) + (1-\alpha(x,\theta)) \delta_{x}(x'),$$
where $\delta_x(x')$ is the Dirac-delta function.
Then the marginal transition kernel satisfies 
$$ K(x' \mid x,\theta) \ge f(x' \mid \theta)  \alpha(x' \mid x,\theta) \ge a \exp(-\ep)$$
since $a(x'\mid x,\theta) \ge \exp(-\ep)$ according to \Cref{prop:acceptance}.
As a result, we have 
\begin{equation}\label{eqn:sufficient-minor}
p(\theta \mid x) K(x' \mid x,\theta) \ge \frac{a}{b} p(\theta)  \cdot a \exp(-\ep)
\end{equation}
\Cref{eqn:sufficient-minor} is sufficient for a minorization condition $K(x'\mid x) \ge a^2 b^{-1} \exp(-\ep)$ to hold on $x' \in \mathbb{X}$ since $p(\theta)$ is proper.

To establish a drift condition, 
let $w: \mathbb{X} \to \mathbb{R}_{>0}$ be integrable with $ v = \int w(x)dx < \infty.$
Then we have the conditional expectation 
\begin{align*}
K_X[w(x)] &= \mathbb{E}\left[w(X^{(t+1)}) \mid X^{(t)} = x \right]\\
&= \int w(x') K(x' \mid x,\theta) dx'\\
& = \int \int w(x') K(x' \mid x,\theta) p(\theta \mid x) d\theta dx'\\
&= \int \int w(x') f(x' \mid \theta) \alpha(x' \mid x,\theta) p(\theta \mid x) d\theta dx' + w(x)  \int (1-\alpha(x,\theta)) p(\theta \mid x) d\theta \\
&\le \int \int w(x') f(x' \mid \theta) p(\theta \mid x) d\theta dx' + w(x) \int p(\theta \mid x) d\theta
\end{align*}
Using $f(x \mid \theta) \le b$, we can show that  
\begin{equation}\label{eqn:drift}
K_X[w(x)] \le bv +  w(x),
\end{equation}
which is the drift condition. 
Combining \Cref{eqn:sufficient-minor,eqn:drift}, we invoke Theorem 8 of~\citet{johnson2013component} to establish geometric ergodicity of the Gibbs sampler. 

When $n \ge 2$, the proof shall proceed by denoting $K(x' \mid x,\theta)$ as the Markov transition kernel on $x,x' \in \mathbb{X}^n$ and similarly for $K(x' \mid x).$ The drift condition becomes $K_X[w(x)] \le b^n v + w(x)$ and minorization condition becomes $K(x' \mid x) \le (a^2b^{-1}\exp(-\ep))^n$.
%
\end{proof}

\section{Log-linear Model: More Details}\label{sec:loglinearExtra}
Our full model, along with conjugate priors is given in the following equation array:
\begin{align}
& \text{prior} & p  & \sim \text{Dirichlet}(\alpha), &\\
    & &  p^k_{i-}  & \sim \text{Dirichlet}(\alpha_i^k) \quad \forall i, & \\
    & \text{data model} & n_{-}  &\sim \text{Multinomial}(N, p_{-}), &\\
    & &  n_{i-}^k  & \sim \text{Multinomial}(n_{i}, p^k_{i-}) \quad \forall i, & \\
      &\text{privacy noise} & L_{ijk} & \iid \text{Laplace}(0, 2K / \epsilon),&\\
    &&  m_{ij}^k & = n_{ij}^k + L_{ijk}\quad \forall i,j,k, &\\
    &\text{privatized output}& s_{dp}&=(m_{ij}^k).&
\end{align}

\section{Linear Regression: More Details and Results}\label{sec:linearRegressionExtra}

\paragraph{Data generating parameters.}
Our experiments use continuous predictors $X_0$, which we model as $X_0^i \iid \mathcal{N}_{p}(m,\Sigma)$. We choose $\Sigma = I_{n}$. We simulate $m_i \iid \mathcal{N}(0,1)$ and hold it fixed at $m = (0.9,-1.17)$.

\paragraph{Conjugate prior distribution.}
Our experiments fix $\sas$ at the data generating value of $\sas = 2.$
Given prior $\beta \sim \mathcal{N}_{p+1}(0, \tau^2 I_{p+1})$ , the posterior distribution $\beta \mid \sas, x,y$ is multivariate Normal 
with covariance matrix $\Sigma_n =  (x^{\top} x/ \sas +  I_{p+1} /\tau^2 )^{-1}$ and mean vector $\mu_n = \Sigma_n (x^{\top}y) / \sas.$ The prior for $\beta$ is $\beta_i \iid \mathcal{N}(0,\tau^2 = 2^2).$

\paragraph{The effect of clamping.} We view clamping as part of the privacy mechanism. The clamping step first truncates $x$ and $y$ values into a fixed range, and then performs data-independent location-scale transformation so that all values of $\tilde{x}$ and $\tilde{y}$ are in the range $[-1,1]$. Although with conjugate priors the confidential data posterior $p(\theta \mid x,y)$ is tractable, the clamped data posterior $p(\theta \mid \tilde{x},\tilde{y})$ no longer enjoys conjugacy and is now intractable. Since the clamping parameters are known, to sample from the clamped data posterior, one can design data-augmentation MCMC algorithms to impute truncated values. Such an imputation algorithm might take $O(n)$ per iteration. We also highlight that as $\epsilon \to \infty$, in which case privacy noise approaches zero, the posterior $p(\theta \mid \sdp)$ approaches $p(\theta \mid \tilde{x},\tilde{y})$.

\begin{figure}[htbp]
\centering
    \includegraphics[width=\linewidth]{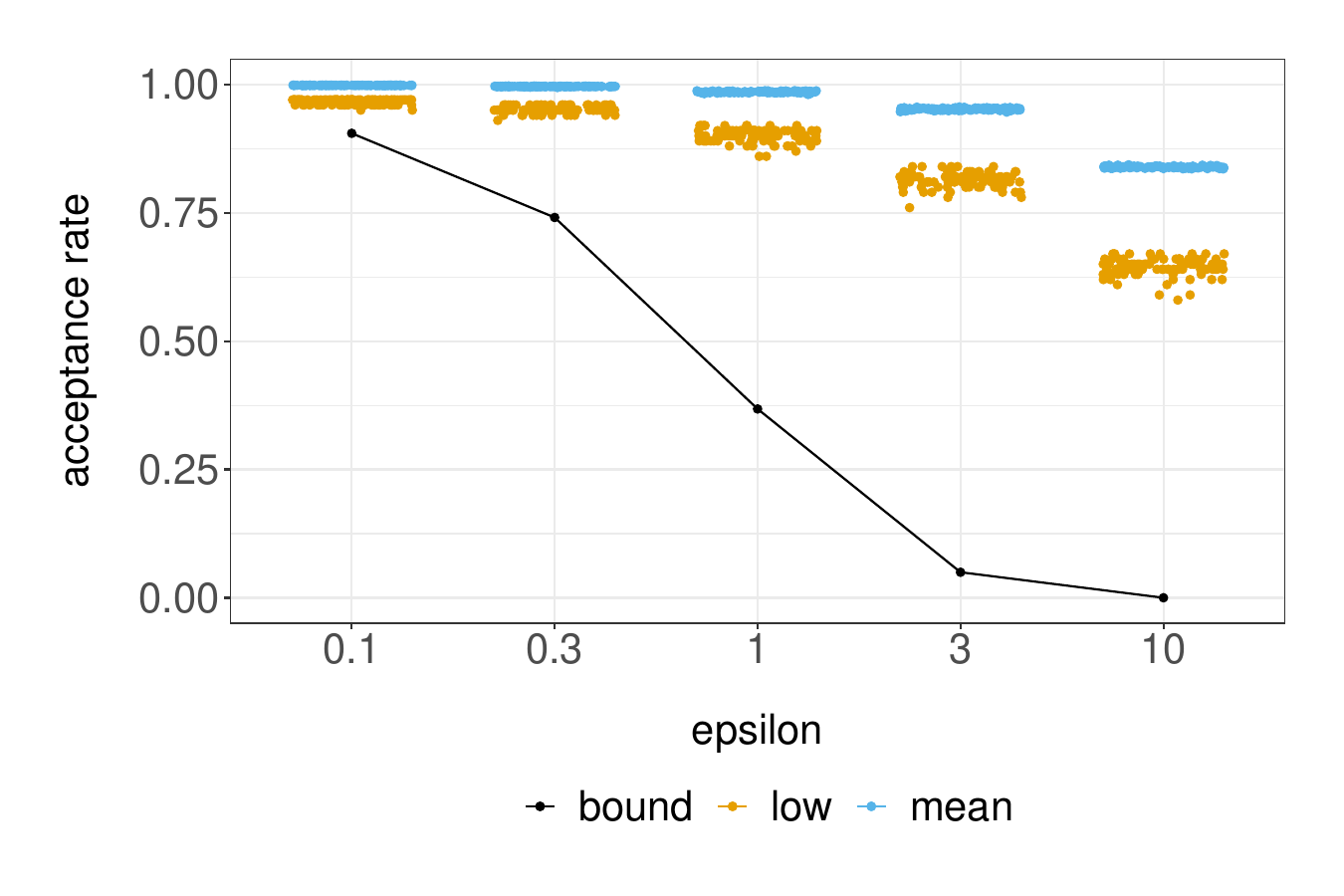}
    \caption{Observed acceptance rates for the log-linear model. The blue (above) point clouds indicate the average acceptance rate, and the orange (below) points indicate the observed minimum acceptance rate of each chain. The solid black line is the lower bound of \Cref{prop:acceptance}.}
    \label{fig:lracceptance.pdf}
\end{figure}

\paragraph{Acceptance rate.}
In \Cref{sec:regression}, 
we report the posterior means of $\beta,\beta_1$ and $\beta_2$ given $\sdp$ 
produced from the same fixed latent database $(x,y)$, with different privacy levels. 
We also report the acceptance rate of $p(x_i \mid x_{-\theta}, \theta, \sdp)$
updates in each iteration of the Gibbs samplers. Recall that for each $\sdp$, 
we run the Gibbs sampler for 10000 iterations and discard the first half for burn-in. 
From \Cref{fig:lracceptance.pdf}, 
we can see that the empirical acceptance rate of the IM proposals is much higher than the lower bound of \Cref{prop:acceptance}.

\paragraph{Posterior credible intervals.}
We repeat the credible interval experiment on log-linear models. 
First we sample one $\beta$ parameter from the prior, and hold this fixed. 
Then for each $\ep$ value, 
we produce 100 confidential databases $(x,y)$ and one private $\sdp$ for each non-private one, and then run a chain for 10,000 iterations targeting $\beta \mid \sdp$. 
After burn-in, from each chain, we produce a $90\%$ credible interval for each $\beta_0,\beta_1$ and $\beta_2$. We then calculate the empirical coverage which is reported in Table \ref{tab:lrcoverage}. 
\begin{table}[htbp]
    \centering
      \[\begin{array}{c|ccc}
            \ep & \beta_0 = -1.79  & \beta_1 = -2.89 & \beta_2  = -0.66 \\\hline
           0.1 & .99 &{\bf.60}& .99\\
            0.3 & 1 &{\bf .66}&.94\\
            1 & 1 & {\bf .84} & {\bf .80}\\
            3 & 1 & {\bf .84}& {\bf .75}\\
            10 & .93 &.87 & .85
        \end{array}\]
        \caption{Coverage of $\beta_0,\beta_1,\beta_2$ in linear regression. Coverage is based on 100 replicates.}
    \label{tab:lrcoverage}
\end{table}

While at $n = 100$, we do not expect the frequentist coverage of the credible
intervals to exactly match the nominal level of .9, note that most of the 
values are close to or above .9. The coverage on $\beta_1$ is lower than 90\%, which might be due to the true parameter being furthest from the prior mean of 0. Another explanation is that data quality loss from truncation and location-scale transformations during the clamping procedure can not be fully recovered by our inference procedure. 

\section{Statement on Computing Resources}
We ran the experiments on an internal cluster. We used a server with a pair of 64-core AMD Epyc 7662 `Rome' processors and with 256GB of RAM. We ran each MCMC chain for 10000 iterations and a typical chain takes approximately 330 seconds for linear regression and approximately 404 seconds for the log-linear model. 


\end{document}